%% file: ms.final.tex
\documentclass{emulateapj}
\usepackage{graphics}
\usepackage{color}
\usepackage{amsmath}
\usepackage{empheq} 
\newcommand{\HI}{H{\sc i}}

\received{}
\revised{}
\accepted{}
\shortauthors{McQuinn et al.}
\shorttitle{Distance to M~51}

\begin{document}
\title{The Distance to M~51}\thanks{Based on observations made with the NASA/ESA Hubble Space Telescope, obtained from the Data Archive at the Space Telescope Science Institute, which is operated by the Association of Universities for Research in Astronomy, Inc., under NASA contract NAS 5-26555.}
\author{Kristen.~B.~W. McQuinn\altaffilmark{1,2}, 
Evan D. Skillman\altaffilmark{2},
Andrew E.~Dolphin\altaffilmark{3},
Danielle Berg\altaffilmark{4}, 
Robert Kennicutt\altaffilmark{5}
}

\altaffiltext{1}{University of Texas at Austin, McDonald Observatory, 2515 Speedway, Stop C1400 Austin, Texas 78712, USA \ {\it kmcquinn@astro.as.utexas.edu}}
\altaffiltext{2}{Minnesota Institute for Astrophysics, School of Physics and Astronomy, 116 Church Street, S.E., University of Minnesota, Minneapolis, MN 55455, USA} 
\altaffiltext{3}{Raytheon Company, 1151 E. Hermans Road, Tucson, AZ 85756, USA}
\altaffiltext{4}{Center for Gravitation, Cosmology and Astrophysics, Department of Physics, University of Wisconsin Milwaukee, 1900 East Kenwood Boulevard, Milwaukee, WI 53211, USA}
\altaffiltext{5}{Institute for Astronomy, University of Cambridge, Madingley Road, Cambridge CB3 0HA, England }

\begin{abstract}
Great investments of observing time have been dedicated to the study of nearby spiral galaxies with diverse goals ranging from understanding the star formation process to characterizing their dark matter distributions. Accurate distances are fundamental to interpreting observations of these galaxies, yet many of the best studied nearby galaxies have distances based on methods with relatively large uncertainties. We have started a program to derive accurate distances to these galaxies. Here we measure the distance to M~51 - the Whirlpool galaxy - from newly obtained Hubble Space Telescope optical imaging using the tip of the red giant branch method. We measure the distance modulus to be $8.58\pm0.10$ Mpc (statistical), corresponding to a distance modulus of 29.67 $\pm$ 0.02 mag. Our distance is an improvement over previous results as we use a well-calibrated, stable distance indicator, precision photometry in a optimally selected field of view, and a Bayesian Maximum Likelihood technique that reduces measurement uncertainties. 

\end{abstract} 


\section{Introduction}\label{sec:intro}
\subsection{Distance is a Fundamental Parameter}

The Spitzer Infrared Nearby Galaxies Survey \citep[SINGS;][]{Kennicutt2003} and its many offspring programs have fundamentally changed what we know about nearby galaxies. In addition to the SINGS infrared observations from the Spitzer Space Telescope, multi-wavelength datasets at unprecedented sensitivity are now available from the GALEX Space Telescope \citep[NGS;][]{GildePaz2007}, the VLA \citep[THINGS;][]{Walter2008}, IRAM 30-m telescope \citep[HERACLES;][]{Leroy2009}, Herschel Space Observatory \citep[KINGFISH][]{Kennicutt2011}, optical integral field spectroscopy \citep[PINGS;][]{Rosales2010}, and optical slit-let spectroscopy \citep[CHAOS;][]{Berg2015}. These high quality, spatially resolved studies have delivered significant advances in areas such as calibrating star formation (SF) rates, modeling stellar radiative transfer in dust, producing unbiased studies of massive SF, mass distributions, and chemical abundance gradients \citep[e.g.,][]{Calzetti2005, Dale2005, Draine2007, Kennicutt2007, Bigiel2008, Leroy2008, Moustakas2010}. 

Surprisingly, secure distance measurements are lacking for many of the SINGS spiral galaxies, including the famous spiral galaxies the Whirlpool (M51; NGC~5194), the Sombrero (M104; NGC~4594), the Sunflower (M63; NGC~5055), and M74 (NGC~628; the archetype grand-design spiral). In the case of M51, measured distances range from 4.9 $-$ 12.2 Mpc from the Tully-Fisher (TF) relation, 6.02 $-$ 8.99 Mpc from type II supernova (SNII), 7.62 $-$ 8.4 Mpc from the planetary nebula luminosity function (PNLF) method, 7.31 $-$ 7.83 Mpc from surface brightness fluctuations (SBF), with additional single measurements from other techniques reported in the literature (see discussion and references in Section~\ref{sec:comparison} and the Appendix). Thus, fundamental results including all luminosity based SF rates, gas masses, physical sizes, radially dependent analysis, spiral structure, and rotation curves, and many other physical properties for M51 rely on $approximate$ distances. This also means that comparisons with other galaxies are prone to systematic offsets.

\begin{figure*}
\includegraphics[width=\textwidth]{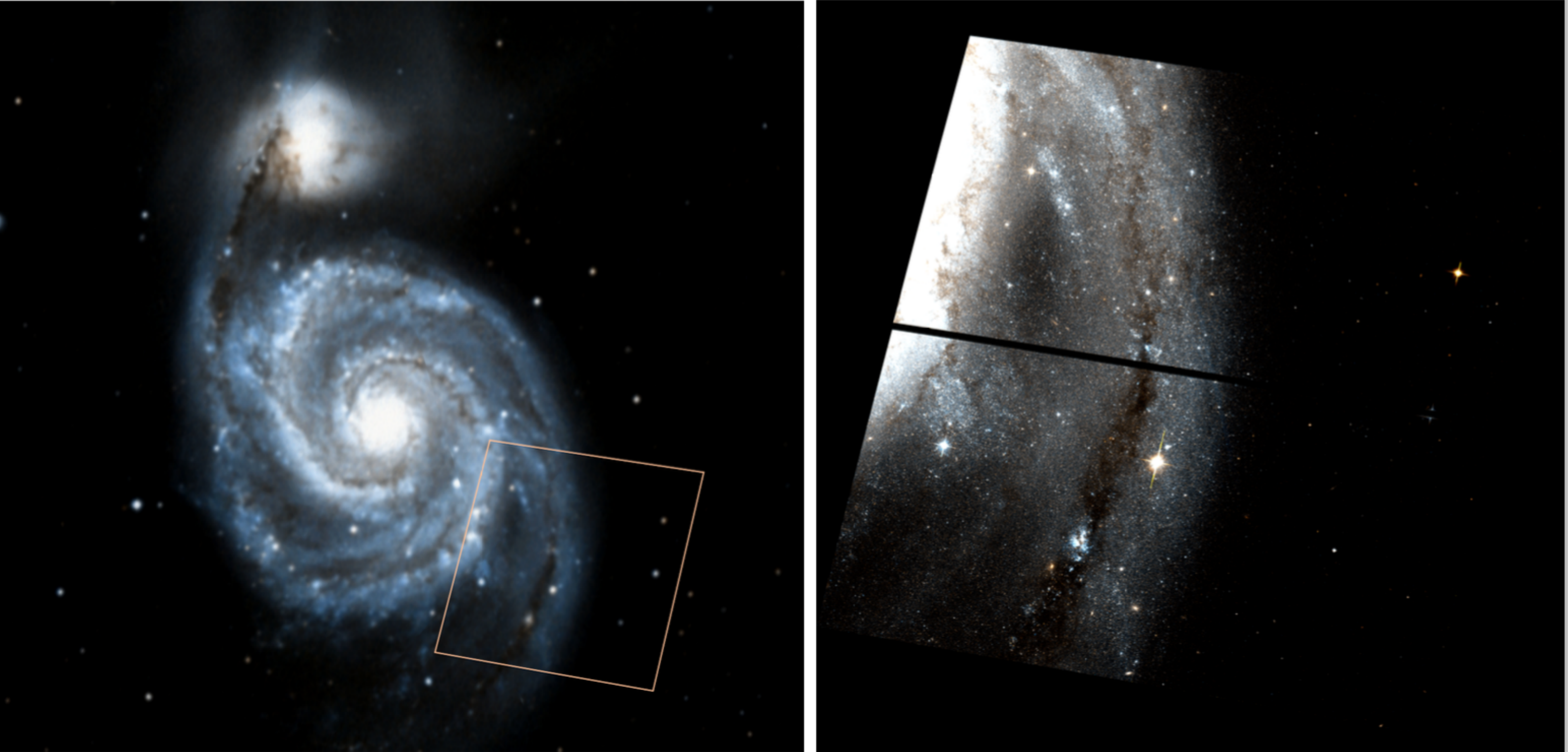}
\caption{{\textit Left:} DSS image of M~51 with the footprint of the HST field of view overlaid. {\textit Right:} HST ACS imaging of the field from combining F606W (blue), F814W (red), and an average of the two filters (green). The images are oriented with North up and East left}
\label{fig:image}
\end{figure*}

\subsection{The Tip of the Red Giant Branch Distance Indicator}
The luminosity of the tip of the red giant branch (TRGB) is arguably one of the most accurate distance indicators to galaxies in the nearby universe. The standard candle TRGB distance methodology arises from the stable and predictable I band luminosity of low-mass stars just prior to the helium flash \citep{Mould1986, Freedman1988, DaCosta1990, Lee1993}. Briefly, low-mass stars ascend the red giant branch (RGB) with a hydrogen burning shell and a convective outer envelope. The shell burning and expansion of the envelope causes an increase in luminosity and a reddening in color, which continues until the core becomes electron degenerate and helium burning is initiated in the core of the star. The core helium burning phase begins abruptly, hydrogen shell burning abates and the luminosity and color of the star is quickly shifted to become fainter and bluer. This `He-flash' is reached at a predictable luminosity, independent of initial stellar mass, with only modest dependencies on stellar metallicity in the I band \citep[e.g.,][]{Lee1993, Salaris1997} enabling the use of the TRGB luminosity as a standard candle for distance determinations.  Indeed, it has been suggested that TRGB distances are preferable to Cepheid distances as the Cepheid period-luminosity relation may not be unique \citep[e.g.,][]{Tammann2008, Mould2008, Ngeow2012}.
 
The TRGB luminosity is well-calibrated to the Hubble Space Telescope (HST) filters, including corrections for the known dependency on metallicity \citep{Rizzi2007a}. As a discontinuity in the I band luminosity function (LF) identifies the TRGB luminosity, the TRGB detection method requires observations of resolved stellar populations in V and I filters that reach $>1$ mag below the TRGB in the I band data. For galaxies within the Local Volume, this means single orbit HST observations (or shorter depending on the approximate distance) obtained in two optical filters can be used to efficiently measure high-precision distances. Because of its combined relative ease and high precision, this approach has been used to accurately determine the distribution of galaxies within our nearby universe for a significant number of galaxies \citep[e.g.,][]{Sakai1997, Harris1998, Cioni2000, Karachentsev2003, McConnachie2004, Dalcanton2009, Tully2009, Jacobs2009}, many of which can be found in the Extragalactic Distance Database \citep[i.e., EDD][]{Tully2009}. 

Here, we present the distance measurements to M~51 based on newly obtained HST optical imaging and the TRGB standard candle method. Our current program includes eight of the SINGS galaxies (M51, M74, M104, M63, NGC~1291, NGC~4559, NGC~4625, NGC~5398); subsequent papers will include results on the remainder of the sample. The paper is organized as follows. \S\ref{sec:obs} describes the observations and data processing. \S\ref{sec:distances} describes the distance determination methods and results. \S\ref{sec:comparison} compares our TRGB distance to previous distance estimates, with expanded discussion of the myriad methods in the Appendix. \S\ref{sec:conclusions} summarizes our conclusions. 

\input{tab1}

\begin{figure*}
\includegraphics[width=\linewidth]{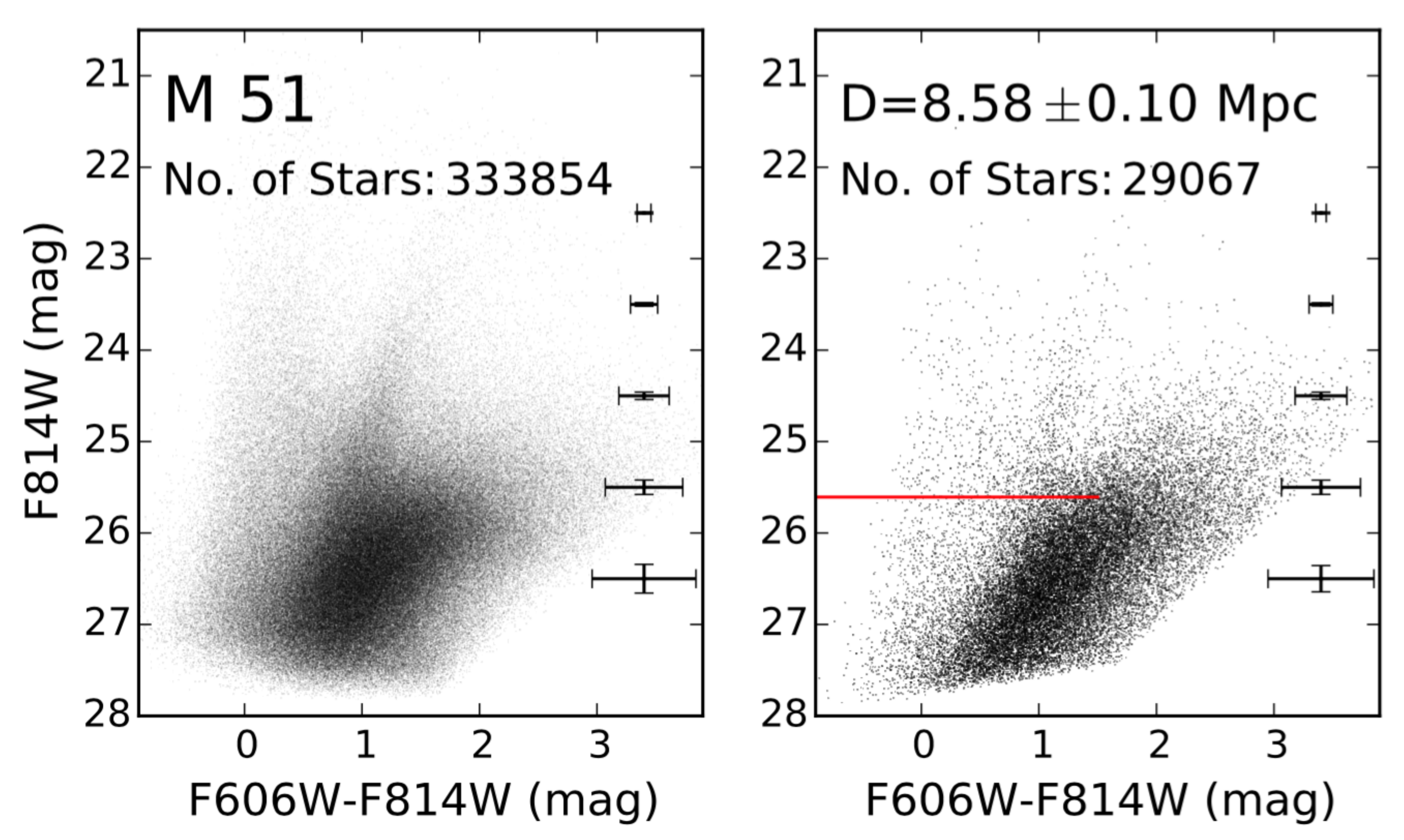}
\caption{{\em Left:} CMD of the full HST field of view. {\em Right:} CMD of the field of view selected for TRGB analysis. Both CMDs have been corrected for foreground extinction. The uncertainties include uncertainties from both the photometry and artificial star tests. The uncertainties in color are larger due to the inclusion of F606W photometry with low SNR. The measured TRGB is marked with an horizontal red line. The photometry in the right panel was transformed using the color-based calibration correction for metallicity. By applying this correction before fitting for the TRGB (instead of applying the correction in the final calibration), the curvature in the RGB is reduced allowing for the TRGB to be measured with a higher degree of certainty.}
\label{fig:cmd}
\end{figure*}

\section{Observations and Photometry}\label{sec:obs}
Table~\ref{tab:properties} lists the coordinates, basic properties, and observation details for M~51. The observations were obtained with the HST using the Advanced Camera for Surveys (ACS) Wide Field Channel (WFC) \citep{Ford1998} as part of the HST-GO-13804 program (PI: McQuinn). Imaging was taken in the F606W and F814W filters during two HST orbits with a 2-point hot pixel dither pattern. Integration times were $\sim$ 2500 s in each filter. The images were processed by the standard ACS pipeline including correcting for the effects of charge transfer efficiency (CTE) non-linearities caused by space radiation damage on the ACS instrument \citep[e.g.,][]{Anderson2010, Massey2010}. 

Because the ACS field of view (FOV) is small relative to the angular extent of M~51, we carefully selected a target field in M~51 to be at a large enough radius from the center of the galaxy such that crowding is minimized, but at a small enough radius that a large number of stars are still observed. Selecting a field at larger radii has additional advantages as it helps reduce the number of asymptotic giant branch (AGB) stars that can mask the TRGB, causing the distance of the galaxy to be underestimated. Internal reddening is also minimized resulting in higher fidelity photometry. Finally, the metallicity of the older RGB stars is lower ensuring that a TRGB F606W$-$F814W color of $\sim1.5$ mag and proper photometric depth in the F606W filter is achieved.

To identify such an optimal field location to observe along the major axis, we used simulations from \citet{Olsen2003} to calculate the surface brightness below which photometric errors would be less than 0.1 mag in the TRGB \citep{Munoz-Mateos2009}. We found typical ACS limiting surface brightnesses of $\mu_V$ $\sim 22.7-23.3$ mag arcsec$^{-2}$ for galaxies at D$=7-9$ Mpc. These limits in surface brightness are expected to yield of order a few tens of thousand stars per ACS FOV at our exposure time, consistent with previous results with ACS observations reaching $\sim1-2$ mag below the TRGB \citep{Makarov2006}. The galactic radius in M~51 corresponding to these surface brightness limits was determined using radial surface brightness profiles from the SINGS survey \citep{Munoz-Mateos2009}. We chose a position angle such that the WFC3 camera could obtain UV observations of central parts of M~51 in parallel.  These parallel observations will be presented elsewhere. 

The left panel of Figure~\ref{fig:image} is a DSS image of the M~51 system, with the ACS field of view overlaid in the bottom right. We chose the field conservatively so that it included not only regions with the surface brightness constraint, but also areas with a range of surface brightness and galactic structure to ensure we could optimize the TRGB measurement. The right panel of Figure~\ref{fig:image} presents a color image combining the F606W (blue), the average of the F606W and F814W images (green), and F814W (red) observations. The images were made using the CTE corrected images (flc.fits files) for each filter and combined with \textsc{Astrodrizzle} from DrizzlePac 2.0. The selected field includes part of an inner spiral arm with higher surface brightness and significant crowding, an outer spiral arm with a range of surface brightness and less crowding, and a region outside the spiral arms. This range in properties allows different regions to be explored to select the stellar populations best-suited to a TRGB distance measurement. 

Photometry was performed on the pipeline processed, CTE corrected files (flc.fits files) using DOLPHOT\footnote{URL: http://americano.dolphinsim.com/dolphot/}, a modified version of HSTphot optimized for the ACS instrument \citep[][]{Dolphin2000}. The resulting photometry list was culled to include high fidelity sources based on a number of measured parameters for each point source including cuts based on the sharpness and crowding parameters. Sharpness indicates whether the flux of a point source is highly peaked or sharp (indicative of a cosmic ray) or too broad (indicative of a background galaxy). Crowding measures how much brighter a star would be if nearby stars had not been fit simultaneously. We filter out stars with higher values of crowding as these sources have higher photometric uncertainties. Specifically, we rejected point sources with (V$_{sharp}+$I$_{sharp}$)$^2>0.075$ and (V$_{crowd}+$I$_{crowd}$)$>0.8$. 

We also filtered the photometry based on the signal-to-noise ratios (SNR) of sources in each filter. Since the TRGB measurement is based on the F814W magnitudes, we applied a minimum SNR of 5$\sigma$, ensuring photometry used in the distance measurement are of higher significance. A more liberal SNR threshold of $2\sigma$ was applied to the F606W photometry. This lower quality threshold avoids excluding sources with a low SNR in the F606W filter but a higher SNR in the F814W photometry and prevents completeness effects from complicating the TRGB measurement. We discuss this further below. Artificial star tests were performed to measure the completeness limit of the images using the same photometry package and filtered on the same parameters. Final photometry lists were corrected for Galactic extinction based on the dust maps of \citet{Schlegel1998} with updated calibration from \citet{Schlafly2011}; these values are shown in Table~\ref{tab:properties}.

Based on the photometry, we found our field selection strategy returned a high number of stars; the filtered photometry list includes over 300k stars. This is significantly higher than the few 10k stars anticipated from the simulations because the field of view includes a range of structure in M~51 with a range in surface brightness. The left panel in Figure~\ref{fig:cmd} shows the CMD of the extinction corrected photometry from the full field of view. The depth of the photometry approximately corresponds to the 50\% completeness level determined from the artificial star tests. Also shown are representative uncertainties per magnitude determined from the PSF fitting photometry and artificial star tests. The bottom of the CMD is flat out to $F606W - F814W \sim1.5$ mag, before incompleteness in the F606W filter begins to become important. This is due to the inclusion of sources in the F606W photometry with a lower SNR threshold and avoids photometric incompleteness from artificially introducing a break in the LF at fainter magnitudes. The CMD has a well-populated RGB sequence with photometry reaching $\sim2$ mag below the approximate TRGB identifiable by eye. The full FOV includes a large number of point sources in the main sequence, helium burning sequences, and possible AGB stars. Also evident is the curvature of the RGB, reflecting the range in metallicity of the RGB stellar population across the FOV.

As seen in Figure~\ref{fig:image}, the chosen FOV overlaps with the inner and outer spiral arms which have a range in stellar density. Thus, we were able to apply spatial cuts to the photometry to focus on the conditions that minimize uncertainties while still including a high number of stars ($\sim$ 29,000, or roughly 10$\%$ of stars in the full FOV). The final region used for the TRGB determination centers on the outer spiral arm where the photometric completeness in the magnitude range of the TRGB increased to $\sim$95\% from $\sim$80\% in the full field of view as measured from artificial star tests. Selecting stars in the outer region of the galaxy has the added advantage of minimizing the contribution of stars in different evolutionary stages, thereby reducing the Poisson noise of non-RGB stars in the TRGB measurement. In addition, outer RGB populations typically have a smaller metallicity range, further reducing the complexity in identifying the break in the LF of the TRGB.  

As discussed below (see Equation~\ref{eq:trgb}), the calibration of the TRGB magnitude includes a metallicity correction based on the average color of the TRGB. This color-based metallicity correction can be applied to the photometry {\em prior} to fitting for the TRGB, thereby reducing the curvature and width of the RGB and increasing the sharpness in the break of the LF. Alongside the CMD of the full FOV in Figure~\ref{fig:cmd}, the right panel shows the CMD of the extinction corrected photometry from the region used for the TRGB analysis after applying the color-based correction for metallicity. We use these data to fit for the break in the LF corresponding to the TRGB. 

\section{Distance Determination from the TRGB}\label{sec:distances}
\subsection{Identifying the TRGB Luminosity with Precision}
We employed two techniques to measure the luminosity of the TRGB in the F814W photometry. First, we applied a Sobel filter edge detection technique to identify the discontinuity in the luminosity function (LF) \citep{Lee1993, Sakai1996, Sakai1997}. For high fidelity photometry, the accuracy of this approach is limited by the binning of the LF (or the smoothing kernel used on the binned LF). Second, we used a Bayesian Maximum Likelihood (ML) technique, which takes into account the photometric uncertainties and completeness of the photometry. The ML technique determines the TRGB luminosities by fitting the observed distribution of stars with a parametric RGB LF \citep[e.g.,][]{Sandage1979, Mendez2002, Makarov2006}. This is an improvement over the Sobel filter because not only does it avoid binning the LF, but the probability estimation takes into account photometric error distribution and completeness from artificial star tests \citep[see][for a full discussion]{Makarov2006}. We assumed the following form for the theoretical LF used in the ML technique:

\begin{subequations}
\begin{empheq}[left={P = }\empheqlbrace]{alignat=2}
	& 10^{(A*(m - m_{TRGB}) + B)}, & \quad \text{if m - m$_{TRGB} \geq 0$}\\
	& 10^{(C*(m - m_{TRGB}))}, & \quad \text{if m - m$_{TRGB} < 0$}
\end{empheq}
\label{eq:ml_form}
\end{subequations}

\noindent where A is the slope of the RGB with a normal prior of 0.30 and $\sigma=0.07$, C is the slope of the AGB with a normal prior of 0.30 and $\sigma=0.2$, B is the RGB jump, and all three are treated as free parameters. This is the same theoretical LF form used in \citet{Makarov2006}. The range in solutions returning log(P) within 0.5 of the maximum gives the uncertainty (as is the case with a normal distribution). 

Because both techniques rely on identifying the break in the LF of the RGB, including sources of all colors in a CMD can skew the results. Therefore, TRGB analysis is typically restricted to stars with colors consistent with the RGB, limiting the number of non-RGB stars contributing to the LF. We selected stars in $F606W - F814W$ color range of $0.5 - 2.0$. The best-fitting value for the TRGB luminosity from the Sobel filter was $F814W = $25.65$\pm0.5$ and from the ML technique was $F814W = $25.61$\pm0.01$. As the ML technique does not rely on binning and includes photometric uncertainties and completeness, we adopt the TRGB luminosity from the ML technique. The identified TRGB is noted in Figure~\ref{fig:cmd}.

\input{tab2}

\subsection{Calibrations and Calculating the Distance Modulus}
We use the zero-point calibrations specific to the HST ACS filters for the TRGB from \citet{Rizzi2007a}, which includes a correction for the known dependency of the TRGB on metallicity. This dependency is reduced in the I band compared to other optical filters \citep[e.g.,][]{Lee1993}, but still has a measurable effect on the calibration. The average color of the RGB stars at the identified TRGB depends on the metallicity of the stars and can therefore be used to correct for this second order effect. For convenience, we reproduce the final relation:

\begin{equation}
M_{F814W}^{ACS} = -4.06+ 0.20 \cdot [(F606W-F814W) - 1.23] \label{eq:trgb}
\end{equation}

\noindent As noted above, we applied the color-correction to the photometry {\em before} identifying the TRGB. Thus, based on the measured TRGB magnitude of 25.61 $\pm$ 0.01, we apply only the zero-point calibration to calculate a distance modulus of 29.67 $\pm$ 0.02 mag. Our final distance measurement to M~51 is 8.58 $\pm$ 0.10. Uncertainties are based on adding in quadrature the uncertainties from the TRGB zeropoint calibration ($\sigma=$ 0.02), the color-dependent metallicity correction ($\sigma=$ 0.01), and the ML uncertainties calculated from the probability distribution function, which include uncertainties from the photometry and artificial star tests. The final numbers are also listed in Table~\ref{tab:distances}.

Note that our statistical uncertainty is small enough that it is likely less than the systematic uncertainty in the measurement. Previously, distances derived from a single data set using three standard candles methods from resolved stellar populations (i.e., based on the TRGB, horizontal branch stars, and red clump stars) differed based on the calibrations used \citep{Rizzi2007b}. In this study, the luminosity of each of the CMD features was consistent with values from previous studies in the literature, but the conversion to a distance measurement varied depending on the calibration. Additional uncertainties may also affect the precision of TRGB distances, but are yet unquantified. For example, there may be systematic uncertainties for the RGB color-based metallicity correction as suggested by photometric and spectroscopic comparisons \citep{Rizzi2007b}. Analysis of synthetic stellar populations suggest a $0.1-0.2$ mag difference in the TRGB luminosity if the RGB population is predominantly young \citep[compared to the predominantly old RGB population used in the TRGB calibration studies from globular clusters;][]{Salaris2005}. This is less likely to affect our distance measurement as the RGB stars in the outer region of M~51 are predominantly old. From the literature, we have the impression that the systematic uncertainty could be of order 0.05 mag, but do not know of a definitive value, so we do not report one here. 

\section{Comparison with Previous Distances}\label{sec:comparison}
\begin{figure}
\includegraphics[width=\linewidth]{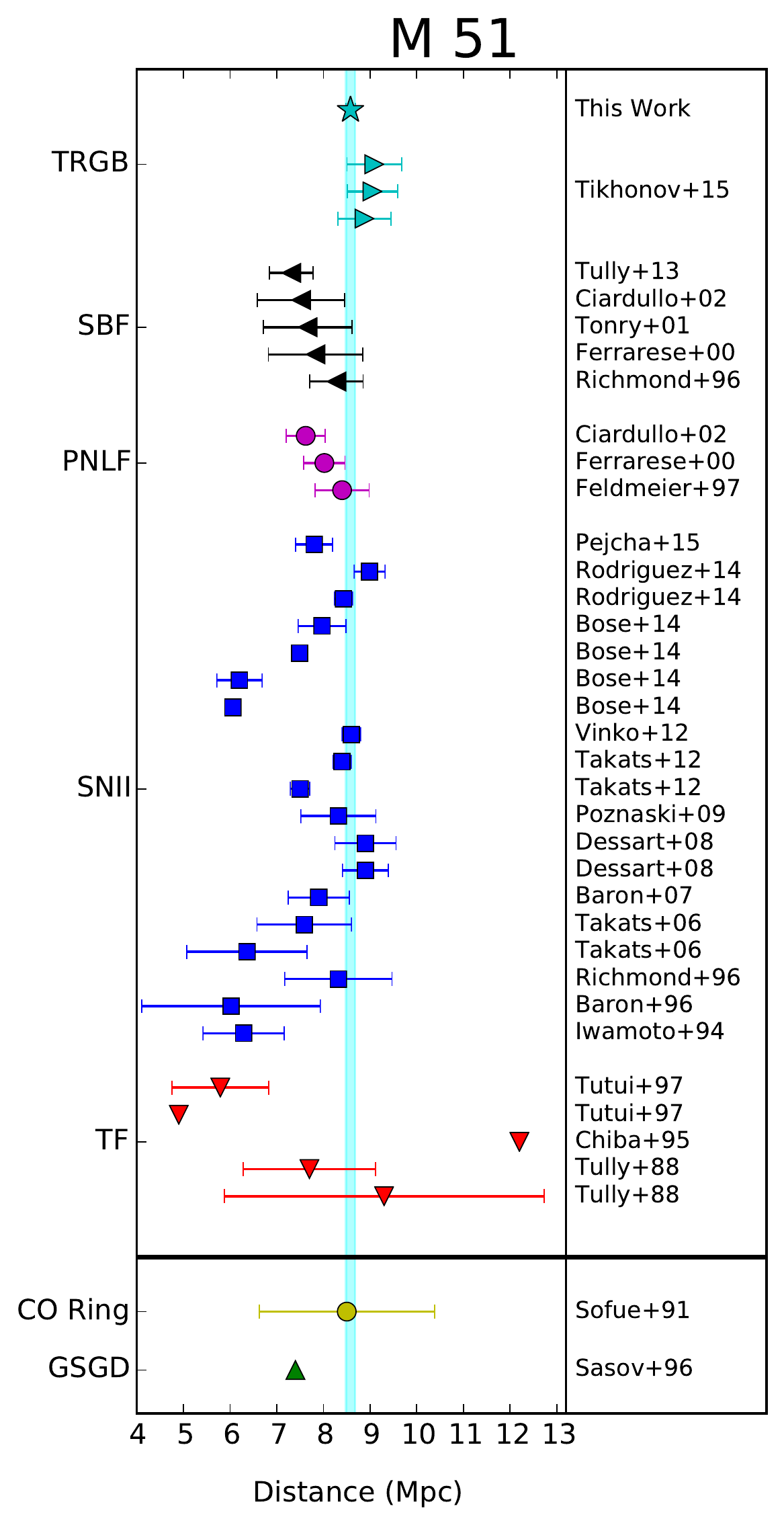}
\caption{Comparison of distance measurements to M~51 from the literature. Our precise TRGB distance measurements lies in the middle of the wide distribution. For ease of comparison with previous results, we also overlay a vertical shaded line in cyan centered on our TRGB measurement whose width encompasses the 1~$\sigma$ uncertainty in our measurement. See Table~\ref{tab:distances} for individual values and references.}
\label{fig:comparison}
\end{figure}

Figure~\ref{fig:comparison} compares our TRGB distance measurement to M~51 with other reported distance measurements using various techniques from the literature. Our distance measurement lies in the middle of the wide distribution. To aid in the comparison with previous results, we overlay a vertical shaded line centered on our distance measurement with a width encompassing the 1$\sigma$ uncertainty in distance. The individual values, methods, and references for the different distances in Figure~\ref{fig:comparison} are listed in Table~\ref{tab:distances}. For the interested reader, we provide more detailed descriptions of each method in the Appendix. Here, we discuss the results from the myriad techniques relative to our new TRGB distance measurement.

In Figure~\ref{fig:comparison}, the various distance measurements are grouped by technique and, within each technique, are listed from the most recent to the oldest. Some of the studies measured multiple distances based on the same data set \citep[i.e.,][]{Tikhonov2015, Rodriguez2014, Bose2014, Takats2012, Dessart2008, Takats2006, Tutui1997, Tully1988}. Typically, each study reported a final number based on combining the individual measurements. We list the separate, individual measurements to show the range found by the different studies, highlighting the challenges in measuring precise distances.

Comparing the results from an individual method across multiple studies, it is readily apparent that results can vary significantly. For example, the distances from Type II SNe span a range from $\sim6$ to $\sim9$ Mpc - a change in 50\%. While the studies of Type II SNe also span nearly two decades of research, two of the most disparate distances (i.e., 6.06 and 8.99 Mpc) were both reported recently in 2014 \citep{Bose2014, Rodriguez2014}. These SNII distances, as well as the others listed in Table~\ref{tab:distances}, are based on the same SNe events, but are calibrated using different assumptions and models (see Appendix). More than 40\% of the SN II distances do not overlap with our TRGB measurement. Despite continued analysis, the distance measurements based on Type II SNe to M~51 do not converge on a value, making interpretation of the different distance values difficult. 

Our TRGB measurement overlaps in uncertainties with the TRGB distances reported recently by \citet{Tikhonov2015} for three different fields in M~51. Our measurement is an improvement over these previous values for a number of reasons. First, the uncertainties on our photometry for the upper RGB are lower as the photometric depth is $\sim1$ mag deeper, reaching $\sim2$ mag below the TRGB. Second, we use a ML technique for our final distance, instead of the Sobel filter used in \citet{Tikhonov2015}, which reduces the uncertainties on our measurement. Finally, we use a TRGB calibration specific to the $HST$ filters. \citet{Tikhonov2015} do not explicitly state the calibration they used, but refer to the \citet{Lee1993} paper when referencing their final TRGB distance modulus. This early paper calibrated the TRGB luminosity in the I-band using RR~Lyrae distances with metallicity corrections based on a $V-I$ color. While similar to the $HST$ calibrations, the difference in filters impacts the conversion from luminosity and color of the TRGB to a final distance measurement. The systematic uncertainty introduced by this different calibration is not quantified in their result.

The SBF method is based on discerning the degradation in resolution of a galaxy with distance. It has typical uncertainties of $\pm1$ Mpc which can vary based on how well a specific galaxy meets the assumption of the SBF approach. For spiral galaxies, particularly interacting galaxies such as M~51 with significant young stellar populations that are highly clustered, the SBF assumption of a smoothly varying older population breaks down. Separate calibrations for spiral galaxies have been attempted \citep{Ferrarese2000}, although this systematic is difficult to take into account and reported SBF distances can therefore underestimate the uncertainties. The older SBF distance measurements overlap with our TRGB distance, but the most recent value does not. 

The PNLF distances overlap with our reported distance for two of the three measurements. Similar to the SBF distances, there are a number of unquantified uncertainties in the PNLF distances including the unknown dependence of the calibration on both the age and the metallicity of the central star in the PN and the extinction corrections on the inherently dusty region around AGB stars in a metal-rich galaxy. 

The TF distances include the original TF distance to M~51 and two additional studies that use modifications to the TF relation. The highest value, 12.2 Mpc, is based on using the traditional TF approach and the \HI\ line width, while the smallest distance, 4.9 Mpc, is based on substituting the CO line width in the relation. Of course, the applicability of the TF relation for an interacting galaxy such as M~51 comes into question as a merger affects emission line widths and inclination corrections. 

There are two measurements from lesser known distance indicators, the ``CO Ring'' method and the ``gravitational stability of the gas disk'' (GSGD) method. These techniques were older attempts to find geometric features that could be predictable and stable in spiral galaxies and, thus, useful as distance indicators. The methods were used for a brief period of time until more efficient and accurate methods were more fully adopted for nearby galaxies (i.e., Cepheid and TRGB distance indicators). They both have higher intrinsic uncertainties; for completeness we include them in the plot. 

Given the large range in distances to M~51 in the literature, one can expect larger uncertainties in the physical characteristics in
the individual studies.  For example, the seminal studies of star formation by \citet{Bigiel2008, Leroy2008} and \citet{Schruba2011} all assumed a distance to M~51 of 8.0 Mpc.  In all cases, the star formation rates, which are proportional to fluxes, are underestimated by 15\%.

\section{Conclusions}\label{sec:conclusions}
We measure the distance to M~51, the Whirlpool galaxy, to be $8.58\pm0.10$ Mpc (statistical). Our measurement is based on $HST$ optical imaging of resolved stellar populations and the TRGB method. The TRGB in the CMD was identified using both a Sobel filter and a ML technique which takes into account photometric uncertainties and incompleteness in the data. The TRGB magnitude was converted to a distance using the calibrations of \citet{Rizzi2007a} specific to the $HST$ filters with a metallicity correction. 

Our reported distance lies in the middle of a wide range of distances ($4.9-12.2$ Mpc) previously reported for M~51. We have employed a number of improvements over these previous measurements including the use of (i) a well-understood distance indicator that is a stable and predictable standard candle, (ii) precision calibration of the method that includes second order corrections, (iii) precise photometry in an uncrowded field with careful application of both spatial and color cuts to the data, and (iv) the ML technique which includes measurements of photometric incompleteness and reduces uncertainty over previous TRGB approaches, such as a Sobel filter, as it does not rely on binning the LF. 

This is the first in a series of papers to measure precise TRGB distances to well-studied spiral galaxies within the Local Volume that do not have secure distances, including M~104, M~74, M~63, NGC~1291, NGC~4559, NGC~4625, and NGC~5398. As described in the Introduction, all galaxies are part of the SINGS, KINGFISH, THINGS, NGS, HERACLES, and PINGS programs and four overlap with the CHAOS programs; many of the results of these programs depend on accurate distances. 

\section{Acknowledgments}
Support for this work was provided by NASA through grant GO-13804 from the Space Telescope Institute, which is operated by Aura, Inc., under NASA contract NAS5-26555. This research made use of NASA's Astrophysical Data System and the NASA/IPAC Extragalactic Database (NED) which is operated by the Jet Propulsion Laboratory, California Institute of Technology, under contract with the National Aeronautics and Space Administration. 

{\it Facilities:} \facility{Hubble Space Telescope}

\appendix
\section{Descriptions of Distance Indicators}\label{sec:appendix}
Here we give brief descriptions of the different methods that have been used to measure the distance to M~51. Individual distance values and references are listed in Table~\ref{tab:distances} and shown in Figure~\ref{fig:comparison}. We have included all publications found under the distance measurements in the NASA/IPAC Extragalactic Database (NED) for M~51. 

\subsection{Surface Brightness Fluctuations (SBF)}
SBF is a secondary distance indicator that uses the change in the linear resolution as a function of distance to galaxies as a statistical measurement of distance \citep{Tonry1988, Jacoby1992}. For closer galaxies, there are fewer stars per pixel and larger variations in the surface brightness across a galaxy, assuming a given galactic structure, than in more distance systems. Of course, SBFs depend on the structure of a galaxy and the ability to both model that structure and to remove extraneous objects (such as globular clusters or contamination) that can introduce non-distance dependent changes in surface brightness. It is most accurate when applied to galaxies with a primarily older stellar population (e.g., ellipticals and the outer regions of spirals).

Because the fluctuations are primarily due to bright RGB stars, increased precision of the SBF calibration depends on observations in the I band which, similar to the TRGB method, are less sensitive to changes in metallicity. The SBF distance to M~51 from the literature using a bluer filter is 8.28 Mpc, while the four measurements using an I band equivalent yield shorter distances ranging from 7.3 $-$ 7.83 Mpc. 

\subsection{Planetary Nebula Luminosity Function (PNLF)}
The luminosity function of planetary nebulae (PNe) is a secondary distance indicator based on determining the high-luminosity cut-off of the [O III] $\lambda$5007 forbidden line measured for a sample of the PNe around post-AGB stars in a single system \citep[e.g.,][]{Jacoby1989, Mendez1993, Feldmeier1997}. Despite theoretical predictions that the brightness of PNe should vary with the age and mass of the central star, these effects are claimed to have a negligible effect on the absolute luminosity of the bright-end PNLF \citep[e.g.][]{Ciardullo1992, Jacoby1999}. Accurate interpretation of PNLFs depend on extinction corrections, which can be difficult in the inherently dusty environments of PNe, particularly in more metal-rich galaxies such as spiral galaxies. The accuracy of the PNLF method is estimated to be a few tenths of a magnitude in distance modulus, although the reported uncertainties listed in Table~\ref{tab:distances} are $\sim0.1$ mag. However, the true uncertainties are likely higher as the unquantified dependence on the age and metallicity of the star and the uncertain extinction corrections  are not taken into account in the PNLF calibration uncertainties. The results on M~51 from the literature vary by $\sim1$ Mpc (see Table~\ref{tab:distances}). 

\subsection{Supernova Type II (SNII)}
The luminosity of core-collapse Type II SNe is widely used as a distance indicator. The distance is determined by correlating the expansion velocity of the supernova measured via spectroscopic data with the change in angular size measured from photometric data. The application of the technique requires the use of an expanding photosphere model (EPM) which assumes that SNII radiate as dilute blackbodies. M~51 has had three recent supernovae, SN1994I, SN2005cs, and SN2011dh.  In Table~\ref{tab:distances} we report nineteen individual distance measurements based on SNe in M~51 ranging from 6.02 $-$ 8.99 Mpc. The differences between values - sometimes for the same SN event - are based on interpretation of the diverse properties of SNe, varying quality of data, the prescription for measuring the expansion velocities, differences in the dilution factor in the EPM, and uncertain extinction corrections. Despite changes in the technique over time, there is no narrowing or convergence of the distance measurements to M~51. Studies reporting distances on the shortest end of the range ($< 7$ Mpc) span 2 decades in time, as do the measurements at the higher end of the reported distance range ($>8.3$ Mpc). In one study, four distances to M~51 are reported; the authors adopt the value that is consistent with distance measurements previously reported using different techniques \citep{Bose2014}. 

\subsection{Tully-Fisher Relation (TF)}
The TF relation is a correlation of two measurements that trace the mass in a spiral galaxy. It relates the \HI\ emission line width (related to the rotation speed and the mass of a galaxy) with the absolute optical luminosity (which also generally traces the mass of a galaxy) \citep{Tully1977}. Because of the relative ease of measuring the \HI\ line width of the gas and the apparent luminosity of a system, it has been a widely used distance indicator. Since its original introduction, the relation has been refined by using measurements that more accurately trace the mass in a spiral, including using circular velocities from \HI\ rotation curves instead of \HI\ line widths and using infrared luminosities instead of optical. Because of the evolving method for determining distances using the basic premise of the TF, different systematic uncertainties make it difficult to compare TF distances in the literature. Furthermore, there are unquantified uncertainties in interpreting the \HI\ line width for interacting galaxies, such as M~51. The four measurements on M~51 using the TF relation differ widely from 7.7$-$9.3 Mpc using a classical approach to the spiral galaxy and its merging companion \citep{Tully1988}, to 4.9 Mpc based on CO line widths, to 12.2 Mpc based on \HI\ linewidths \citep{Tutui1997}, and finally to 5.79 Mpc using a hybrid approach that included central surface brightness measurements and scale length estimates with rotation velocity \citep{Chiba1995}. Because of the disparate methods, care must be taken when using a reported TF distance for any individual galaxy \citep[cf.][]{Freedman2010, Zaritsky2014}. 

\subsection{CO Ring (CO)}
While not widely used, the geometry of molecular rings in spiral galaxies was put forward as a possible distance indicator \citep{Sofue1991}. The approach is based on the assumption that rings of molecular gas will form in spirals within a specific range of linear distance from the galaxy center. The angular scale of an observed ring is then converted to a linear scale by assuming a distance. We found an individual measurement to M~51 using this approach of $8.5\pm1.7$ Mpc. 

\subsection{Gravitational Stability of the Gaseous Disks (GSGD)}
We found one distance measurement in the literature which used the assumption that there is some radius in spiral galaxies where the total gas surface density is proportional to a critical surface density for gravitational stability of a gaseous disk. Using the azimuthally averaged radial distribution of atomic and molecular hydrogen, in conjunction with the measured rotational velocity of the gas, \citet{Zasov1996} calibrated a distance indicator for a sample of spiral galaxies. The results show a correlation between the measured values but with a high degree of scatter. No uncertainties were quantified for the distance of 7.4 Mpc reported to M~51.

\renewcommand\bibname{{References}}
\bibliography{../../bibliography.bib}

\end{document}

%% file: tab1.tex
\begin{table}
\begin{center}
\caption{M~51 Properties and Observations}
\label{tab:properties}
\end{center}
\begin{center}
\vspace{-15pt}
\begin{tabular}{ll}
\hline 
\hline 
Parameter				& Value	\\
\hline
RA (J2000)			& $13:29:52.7$ \\
Dec (J2000)			& $+47:11:43$	\\
$A_{F606W}$			& 0.086 mag \\
$A_{F814W}$			& 0.053 mag \\
F606W exp. time		& 2681 sec \\
F814W exp. time		& 2681 sec \\

\hline
\\
\end{tabular}
\end{center}
\vspace{-10pt}
\tablecomments{Coordinates are for M~51a, the spiral galaxy of the interacting pair making up the M~51 system. Observation times are from program GO$-$13804 (PI McQuinn). Galactic extinction estimates are from \citet{Schlafly2011}.}

\end{table}

%% file: tab2.tex
\begin{table*}
\begin{center}
\caption{Distance Measurements to M~51}
\label{tab:distances}
\end{center}
\begin{center}
\vspace{-15pt}
\begin{tabular}{llll}
\hline 
dm (mag)		& D (Mpc)	& Reference	& Data	\\
\hline 
\hline
\\
29.67$\pm0.02$	& $8.58\pm0.10$ 	& \textbf{This work} 	& new observations \\
\\
\multicolumn{4}{c}{\textbf{Tip of the Red Giant Branch (TRGB)}}\\
29.78	$\pm$0.13	& 9.05	& \citet{Tikhonov2015} & archival \\
29.79	$\pm$0.14	& 9.09	& \citet{Tikhonov2015} & archival \\
29.74	$\pm$0.14	& 8.88	& \citet{Tikhonov2015} & archival \\
\\
\multicolumn{4}{c}{\textbf{Surface Brightness Fluctuations (SBF)}} \\
29.32	$\pm$0.14	& 7.31	& \citet{Tully2013} & \citet{Tonry2001} \\
29.38	$\pm$0.27	& 7.52	& \citet{Ciardullo2002} & \citet{Tonry2001} \\
29.42	$\pm$0.27	& 7.66	& \citet{Tonry2001} & new observations \\
29.47	$\pm$0.28	& 7.83	& \citet{Ferrarese2000} & \citet{Tonry2001} \\
29.59	$\pm$0.15	& 8.28	& \citet{Richmond1996} &  new observations \\
\\
\multicolumn{4}{c}{\textbf{Planetary Nebulae Luminosity Function (PNLF)}} \\
29.41	$\pm$0.12	 & 7.62 	& \citet{Ciardullo2002} & \citet{Feldmeier1997} \\
29.52	$\pm$0.12	 & 8.02	& \citet{Ferrarese2000} & \citet{Feldmeier1997} \\
29.62	$\pm$0.15	 & 8.40	& \citet{Feldmeier1997} & new observations \\
\\
\multicolumn{4}{c}{\textbf{Optical Type II Supernova (SNII)}}\\
29.46	$\pm$0.11		& 7.80	& \citet{Pejcha2015} &  SN 2005cs \\
29.77	$\pm$0.08	& 8.99	& \citet{Rodriguez2014} &  SN 2005cs \\
29.63	$\pm$0.05	& 8.43	& \citet{Rodriguez2014} &  SN 2005cs \\
29.51	$\pm$0.14	& 7.97	& \citet{Bose2014} &  SN 2005cs \\
29.37	$\pm$0.04	& 7.49	& \citet{Bose2014} &  SN 2005cs \\
28.96	$\pm$0.17	& 6.20	& \citet{Bose2014} &  SN 2005cs \\
28.91	$\pm$0.05	& 6.06	& \citet{Bose2014} &  SN 2005cs \\
29.62	$\pm$0.05	& 8.40	& \citet{Vinko2012} &  SN 2005cs \& SN 2011dh\\
29.67	$\pm$0.05	& 8.60	& \citet{Takats2012} &  SN 2005cs \\
29.38	$\pm$0.06	& 7.50	& \citet{Takats2012} &  SN 2005cs \\
29.61	$\pm$0.21	& 8.32	& \citet{Poznanski2009} & SN 2005cs \\
29.75	$\pm$0.12	& 8.90	& \citet{Dessart2008} & SN 2005cs \\
29.75	$\pm$0.16	& 8.90	& \citet{Dessart2008} & SN 2005cs \\
29.50	$\pm$0.18	& 7.90 	& \citet{Baron2007} &  SN 2005cs \\
29.40	$\pm$0.29	& 7.59	& \citet{Takats2006} & SN 2005cs \\
29.02	$\pm$0.44	& 6.36	& \citet{Takats2006} & SN 2005cs \\
29.60	$\pm$0.30	& 8.32	& \citet{Richmond1996} & SN 1994I \\
28.90	$\pm$0.69	& 6.02 	& \citet{Baron1996} & SN 1994I \\
29.20	$\pm$0.30	& 6.29	& \citet{Iwamoto1994} & SN 1994I \\
\\
\multicolumn{4}{c}{\textbf{Tully-Fisher Relation (TF)}}\\
28.45				& 4.90	& \citet{Tutui1997} & archival \\
30.43				& 12.2	& \citet{Tutui1997} & archival \\
28.88	$\pm$0.39	& 5.79	& \citet{Chiba1995} & archival \\
29.43	$\pm$0.40	& 7.70	& \citet{Tully1988} & archival \\
29.85	$\pm$0.80	& 9.30	& \citet{Tully1988} & archival \\
\\
\hline
\multicolumn{4}{c}{\textbf{Gravitational Stability of Gas Disk (GSGD)}}\\
29.35				& 7.40	& \citet{Zasov1996}	& archival \\
\\
\multicolumn{4}{c}{\textbf{CO Ring}}\\
29.65	$\pm$0.48	& 8.50	& \citet{Sofue1991} & archival \\
\hline\hline
\end{tabular}
\end{center}
\vspace{-10pt}
\tablecomments{Distance measurements from the literature from various techniques. The Reference column lists the source of the reported measurement. The Data column lists whether the observations were original to the study, from data archives, or a re-calibration of existing work in the literature. In the case of the SN II distances, the individual SN event(s) used in each study is listed. Figure~\ref{fig:comparison} shows the distribution of the measurements.  Details on each method can be found in the Appendix.}

\end{table*}